\documentstyle[twoside,fleqn,espcrc2]{article}

\newcommand{\AmS}{{\protect\the\textfont2
  A\kern-.1667em\lower.5ex\hbox{M}\kern-.125emS}}
\newcommand{\be}{\begin{equation}}  
\newcommand{\ee}{\end{equation}}
\newcommand{\ba}{\begin{eqnarray}}  
\newcommand{\ea}{\end{eqnarray}}

\newcommand{\bea}{\begin{eqnarray}}
\newcommand{\eea}{\end{eqnarray}}

\def\dfrac#1#2{{\displaystyle {#1 \over #2}}}

\title{Quark masses and the chiral condensate with a 
      non-perturbative renormalization procedure}
\author{V.~Gim\'enez\address{Univ. de Val\`encia and IFIC,
Dr. Moliner 50, E-46100, Burjassot, Val\`encia, Spain},
L.~Giusti\address{Scuola Normale Superiore and INFN Sezione di Pisa, P.zza dei 
         Cavalieri 7 - I-56100 Pisa Italy}\thanks{Talk presented by L.~Giusti}, 
F.~Rapuano\address{Universit\`a di Roma \lq La Sapienza\rq ~and
         INFN, Sezione di Roma, P.le A. Moro 2, I-00185 Roma, Italy},
M.~Talevi\address{Department of Physics and Astronomy, University of 
Edinburgh, Edinburgh EH9 3JZ, UK},
A.~Vladikas\address{INFN-Sezione di Roma II, Universit\`a di Tor Vergata, 
Via della Ricerca Scientifica 1 Roma, Italy}.}
\begin{document}
\begin{abstract}
\vspace{-.3cm}
We determine the quark masses and the chiral condensate in the 
$\overline{MS}$ scheme at NNLO from Lattice QCD in the quenched
approximation at $\beta=6.0$, $\beta=6.2$ and $\beta=6.4$ using both the 
Wilson and the tree-level improved SW-Clover fermion  action. We extract 
these quantities using the Vector and the Axial Ward 
Identities and non-perturbative values of the renormalization constants. 
We compare the results obtained with the two methods and we study the $O(a)$ 
dependence of the quark masses for both actions. 
\end{abstract}
\maketitle
\vspace{-1.cm}
\section{INTRODUCTION}
To describe the light hadron spectrum one considers 
massless QCD a good approximation to this world and the
vacuum not symmetric under chiral transformations. The chiral condensate is
the order parameter which governs the spontaneous chiral symmetry breaking.
The pseudoscalar mesons are identified with the goldstone bosons and their
physical masses are attributed to the mass term in the QCD Hamiltonian. 
Since free quarks are not physical states, quark masses cannot be
measured directly in the experiments and can be determined from the meson 
spectrum using non-perturbative techniques. 
On the lattice one can compute the quark masses 
and the chiral condensate from first principles and it is the only procedure 
that can be systematically improved. The methods and the symbols we have used 
and all the results we have obtained are fully described in 
\cite{finale,finale2}.        
\vspace{-.3 cm}
\section{QUARK MASSES}   
\begin{figure}[t]   
\caption{$Z_S^{RGI}$ vs $\mu$ for the run W64.}
\protect\label{fig.1}
       \setlength{\unitlength}{1truecm}
       \begin{picture}(3.0,3.0)
         {\includegraphics{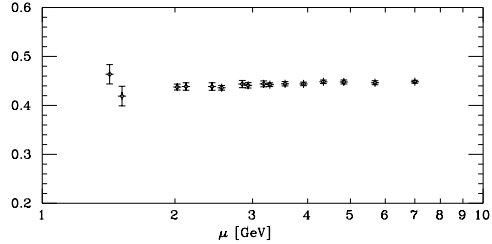}}
       \end{picture}
\end{figure}

\vspace{-.2 cm}
The usual on-shell mass definition cannot be used for quark masses and 
their values depend on the theoretical definition adopted. In the following we
will give our final results for the running quark masses defined in the 
$\overline{MS}$ scheme.
Quark masses can be defined from the Vector Ward Identity (VWI). Neglecting 
terms of $O(a)$, the VWI between on-shell hadronic states can be written as 
\cite{bochi} 
\begin{equation}
\label{eq:vwi}
\langle\alpha|\partial^\mu \tilde V_\mu|\beta\rangle =
\frac{1}{2}\Bigl(\frac{1}{k_2}-\frac{1}{k_1}\Bigr)   
            \langle\alpha| S |\beta\rangle\; .
\end{equation}
Eq.~(\ref{eq:vwi}) fixes the relation  between
the lattice bare quark mass in lattice units and the hopping parameter, i.e.
for the Wilson action $m = 1/2 (1/k - 1/k_c)$.        
\begin{table*}[hbt]
\setlength{\tabcolsep}{.70pc}
\newlength{\digitwidth} \settowidth{\digitwidth}{\rm 0}
\catcode`?=\active \def?{\kern\digitwidth}
\label{tab:latparams}
\caption{Summary of the parameters of the runs analyzed in this work.} 
\begin{tabular}{llllllll}
\hline         
\multicolumn{8}{c}{Matrix Elements} \\
        &C60a&C60b&W60&C62&W62&C64&W64\\
\hline
$\beta$ &6.0 &6.0 &6.0&6.2 &6.2 &6.4&6.4\\
Action  &SW  &SW  &Wil&SW  &Wil &SW &Wil\\
\# Confs&490 &600 &320&250 &250 &400&400\\
Volume  &$18^3\times 64$&$24^3\times 40$&$18^3\times 64$&
         $24^3\times 64$&$24^3\times 64$&
         $24^3\times 64$&$24^3\times 64$\\
\hline
\multicolumn{8}{c}{Renormalization Constants} \\
\multicolumn{1}{c}{ } &C60Z&W60Z&C62Z&W62Z&C64Z&W64Z&\multicolumn{1}{c}{ }\\
\hline
\multicolumn{1}{c}{$\beta$} &6.0&6.0&6.2&6.2&6.4&6.4&\multicolumn{1}{c}{ }\\
\multicolumn{1}{c}{Action } &SW &Wil&SW &Wil&SW &Wil&\multicolumn{1}{c}{ }\\
\multicolumn{1}{c}{\# Confs}&100&100&180&100&60 &60 &\multicolumn{1}{c}{ }\\
\multicolumn{1}{c}{Volume } &$16^3\times 32$&$16^3\times 32$&
                             $16^3\times 32$&$16^3\times 32$&
                             $24^3\times 32$&$24^3\times 32$&
                             \multicolumn{1}{c}{ }\\
\hline
\end{tabular}
\end{table*}
Quark masses can also be extracted from the Axial Ward Identity (AWI). 
Neglecting terms of $O(a)$, the AWI can be
written as \cite{bochi}
\begin{equation}
Z_A \langle\alpha|\partial^\mu A_\mu^a|\beta\rangle = 2 (m_0 - \overline{m}) 
\langle\alpha| P^a|\beta\rangle\; , 
\end{equation}
where $\overline{m}$ is defined in \cite{bochi}.
The light and strange quark masses are determined by fixing to their experimental
values the masses of the $\pi$ and $K$ mesons.   
The standard perturbative approach \cite{finale,kenway} uses lattice and the continuum 
perturbation theory to connect the ``bare'' lattice quark mass to 
the renormalized ${\overline{MS}}$ one.
\setlength{\tabcolsep}{.35pc}
\begin{table}
\caption{Quark Masses an the Chiral Condensate from the VWI in MeV. 
$\overline{MS}$ values are at $\mu=2$ GeV.}
\label{tab:qmas_spect}
\begin{tabular}{llll}
\hline
 Run     &$m_l^{\overline{MS}}$  
         &$m_s^{\overline{MS}}$
         &$- \dfrac{1}{N_f} \langle \bar \psi \psi \rangle_1$ \\
\hline
W60    & $5.8(2)(1)$ & $130(3)(2)$ & $(247 \pm 2 \pm 1)^3$  \\
W62    & $5.4(2)(1)$ & $124(4)(2)$ & $(250 \pm 3 \pm 1)^3$  \\
W64    & $4.9(2)(1)$ & $112(5)(2)$ & $(258 \pm 4 \pm 1)^3$  \\
\hline
\end{tabular}
\vspace{-.3 cm}
\end{table}
The scale $1/a$, where $a$ is the lattice spacing, of our 
simulations is $a^{-1}\simeq 2-4$ GeV. At these scales we expect small 
non-perturbative effects. However the
``tadpole'' diagrams, which are present in lattice perturbation
theory, can give rise to large perturbative corrections and hence to large
uncertainties in the 1-loop matching procedure at values of 
$\beta=6/g_L^2=6.0-6.4$.
A non-perturbative renormalization (NP) technique eliminate these 
uncertainties \cite{rinnp,luescher}. $Z_A$, $Z_P$ and $Z_S$ can be 
calculated by imposing the renormalization conditions, proposed in 
\cite{rinnp}, on the 
quark states of momentum $p^2=\mu^2$ and in the Landau gauge \cite{finale2}.
This procedure works if $\mu$ satisfies the condition 
$\Lambda_{QCD}\ll \mu \ll 1/a$ to avoid chiral symmetry breaking effects, 
large higher-order perturbative corrections
and discretization errors. In figure~\ref{fig.1} we show the scalar
renormalization constant divided by its Renormalization Group (RG) evolution 
($Z_S^{RGI}$). The data show that discretization errors are within 
statistical errors in the range $0.5 < \mu a < 2$  where the 
NP renormalization is applied to compute the quark masses. 
The NP method allows a fully non perturbative definition of the
renormalized quark masses in the $RI$ scheme
\ba
m^{RI}(\mu) & = & \frac{1}{Z_S^{RI}(\mu a)} m a^{-1}\nonumber\\
m^{RI}(\mu) & = & \frac{Z_A^{RI}}{Z_P^{RI}(\mu a)}\rho a^{-1}\; 
\ea 
where for large time separations
\begin{equation}
\rho(a) = \frac{1}{2}\sinh(M_{PS})\frac{\langle A_0(\tau) P(0) \rangle}
     {\langle P(\tau) P(0)\rangle}\; .
\end{equation}
The $\overline{MS}$ definition of the renormalized quark masses is 
intrinsically perturbative and can be related to the $RI$ one through 
continuum perturbation theory only: 
\be
m^{\overline{MS}}(\mu)=  U_m^{\overline{MS}}(\mu,\mu')
\frac{Z_m^{\overline{MS}}(\mu')}{Z_m^{RI}(\mu')} m^{RI}(\mu')\; ,
\ee
where $U_m^{\overline{MS}}(\mu,\mu')$ is the RG
evolution of the quark mass. $Z_m^{\overline{MS}}/Z_m^{RI}$ is the  
matching factor computed in perturbation theory at scales 
$\mu\simeq 2-4$ GeV large enough to 
avoid non-perturbative effects and/or higher order corrections.
\setlength{\tabcolsep}{.35pc}
\begin{table}
\caption{Quark Masses and the Chiral Condensate from the AWI 
in MeV. $\overline{MS}$ values are at $\mu=2$ GeV.}                                  
\label{tab:qmas_wi}
\begin{tabular}{llll}
\hline
 Run     &$m_l^{\overline{MS}}$  
         &$m_s^{\overline{MS}}$
         &$ - \dfrac{1}{N_f} \langle \bar \psi \psi \rangle_2$ \\
\hline
C60a   & $6.0(2)(9)$ & $136(4)(20)$ & $(242 \pm 3 \pm 12)^3$  \\
C60b   & $5.7(2)(8)$ & $132(4)(19)$ & $(244 \pm 2 \pm 12)^3$  \\
W60    & $5.7(2)(8)$ & $127(4)(17)$ & $(248 \pm 2 \pm 11)^3$  \\
C62    & $5.8(5)(6)$ & $131(7)(14)$ & $(245 \pm 4 \pm 9 )^3$  \\
W62    & $5.4(2)(5)$ & $122(4)(12)$ & $(251 \pm 3 \pm 8 )^3$  \\
C64    & $4.4(3)(3)$ & $104(5)(6)$  & $(265 \pm 4 \pm 5 )^3$  \\
W64    & $4.7(2)(3)$ & $108(5)(8)$  & $(262 \pm 4 \pm 6 )^3$  \\
\hline
\end{tabular}
\end{table}
\vspace{-.3 cm}
\section{THE CHIRAL CONDENSATE}   
The chiral condensate can be defined using the AWI arising from the variation
of the non-singlet pseudoscalar density. The integrated AWI becomes
\cite{finale2,bochi}
\[
\dfrac{1}{N_f} \langle \bar \psi \psi \rangle =
\lim_{m_0 \rightarrow m_C}
2 (m_0 - \overline m) \int d^4 x \langle P(x) P(0) \rangle 
\; .
\]
In the chiral limit one can show that the above expression is equivalent to 
\be
\label{eq:wiaa}
\dfrac{1}{N_f} \langle \bar \psi \psi \rangle = 
- \lim_{m_0 \rightarrow m_C} \dfrac{f_P^2 M_P^2}
{4 (m_0 - \overline m)}\; ,
\ee
which is the familiar Gell-Mann--Oakes--Renner (GMOR) relation.  
We write the relation (\ref{eq:wiaa}) for the renormalized condensate as 
\ba
\dfrac{1}{N_f} \langle \bar \psi \psi \rangle_1
& = & - \dfrac{1}{2} a^{-1} f_\chi^2 Z_S C^{HS} \nonumber\\
\dfrac{1}{N_f} \langle \bar \psi \psi \rangle_2
& = & - \dfrac{1}{2} a^{-1} f_\chi^2 \dfrac{Z_P}{Z_A} C^{AWI}\; ,
\ea 
where $f_\chi=0.1282$~GeV is the ``experimental'' value in physical units
\cite{finale2} and 
\ba
M^2_{P} & = & C^{HS} \left( \dfrac{1}{\kappa} - \dfrac{1}{\kappa_C} \right)
\nonumber\\ 
2 a\rho  & =  & \dfrac{1}{C^{AWI}} M^2_{P} \; .
\ea 
The main advantages of the above formulas are to avoid the error 
amplification of the standard method because we are left with only one power of
the UV cutoff $a^{-1}$ and to determine the slope $C^{HS}$
($C^{AWI}$) without extrapolation to the chiral limit. 
Note that since we work at $\kappa$ values typical of the 
strange quark mass, we are implicitly assuming that the slope will not change 
in the chiral region. 
\vspace{-.3 cm}     
\section{RESULTS}
We have computed the renormalization constants of bilinear quark operators with
the NP method proposed in \cite{rinnp}. The $\mu$-dependence of $Z_A$
and $Z_S$ are in excellent agreement with the RG predictions. $Z_P$ is in good
agreement but the chiral symmetry
breaking effects deserve a more accurate analysis. 
The parameters and the action used in each simulation are listed in Table 1 
and the main results we have obtained are reported in Tables \ref{tab:qmas_spect} and
\ref{tab:qmas_wi}. The data at $\beta=6.4$ have to be consider for an 
exploratory study only, since the physical volume and the time extension of the lattice 
are too small to be considered reliable. The $\overline{MS}$ values of the 
quark masses and the chiral condensate
reported in Tables \ref{tab:qmas_spect} and \ref{tab:qmas_wi} show a very good
agreement between the values extracted from the
VWI and from the AWI using the non perturbative
determinations of the renormalization constants, while the same comparison
using the perturbative values of $Z_A$, $Z_P$ and $Z_S$ fails giving
inconsistent results for the two methods \cite{finale}. This pattern is also
found by \cite{mescia,staggered} for different fermion actions.  
In the $\beta$ range studied, there is {\it no}
statistical evidence for an "$a$" dependence of the quark masses and the chiral
condensate. 
We believe that the best estimates for the light and 
strange quark masses and the chiral condensate are 
$m_l^{\overline{MS}}(2\; GeV)=( 5.7 \pm 5 \pm 8 \pm 8 )\; MeV$,
$m_s^{\overline{MS}}(2\; GeV)=( 130 \pm 8 \pm 15 \pm 15)\; MeV$ and 
$- \dfrac{1}{N_f} \langle \bar \psi \psi \rangle^{\overline{MS}}(2\; GeV)
=( 245 \pm 4 \pm 9 \pm 7 \; MeV)^3$, where the first error is 
statistical, the second is due to the non-perturbative renormalization and 
the third is an estimate of the overall systematic errors on the bare
quantities \cite{finale}.    

\end{document}